\documentclass[%
reprint,
superscriptaddress,
 amsmath,amssymb,
 aps,
prl,
]{revtex4-2}

\usepackage{graphicx}% Include figure files
\usepackage{dcolumn}% Align table columns on decimal point
\usepackage{bm}
\usepackage{float}% bold math
\begin{document}

%\preprint{APS/123-QED}

\title{Transverse resistance due to electronic inhomogeneities in superconductors}

\author{Shamashis Sengupta}
\email[]{shamashis.sengupta@ijclab.in2p3.fr}
\affiliation{Universit\'{e} Paris-Saclay, CNRS/IN2P3, IJCLab, 91405 Orsay, France}

\author{Alireza Farhadizadeh}
\email[]{alireza.farhadizadeh@liu.se}
\affiliation{Department of Physics, Chemistry and Biology (IFM), Link\"oping University, Link\"oping SE-581 83, Sweden}

\author{Joe Youssef}
\affiliation{Universit\'{e} Paris-Saclay, CNRS/IN2P3, IJCLab, 91405 Orsay, France}

\author{Sara Loucif}
\affiliation{Universit\'{e} Paris-Saclay, CNRS/IN2P3, IJCLab, 91405 Orsay, France}

\author{Florian Pallier}
\affiliation{Universit\'{e} Paris-Saclay, CNRS/IN2P3, IJCLab, 91405 Orsay, France}

\author{Louis Dumoulin}
\affiliation{Universit\'{e} Paris-Saclay, CNRS/IN2P3, IJCLab, 91405 Orsay, France}

\author{Kasturi Saha}
\affiliation{Department of Electrical Engineering, Indian Institute of Technology Bombay, Mumbai 400076, India}

\author{Sumiran Pujari}
\affiliation{Department of Physics, Indian Institute of Technology Bombay, Mumbai 400076, India}

\author{Magnus \'Oden}
\affiliation{Department of Physics, Chemistry and Biology (IFM), Link\"oping University, Link\"oping SE-581 83, Sweden}

\author{Claire Marrache-Kikuchi}
\affiliation{Universit\'{e} Paris-Saclay, CNRS/IN2P3, IJCLab, 91405 Orsay, France}

\author{Miguel Monteverde}
\email[]{miguel.monteverde@universite-paris-saclay.fr}
\affiliation{Universit\'e Paris-Saclay, CNRS, Laboratoire de Physique des Solides, 91405, Orsay, France}

\begin{abstract}
Phase transitions in many-body systems are often associated with the emergence of spatial inhomogeneities. Such features may develop at microscopic lengthscales and are not necessarily evident in measurements of macroscopic quantities. In this work, we address the topic of distribution of current paths in  superconducting films. Typical lengthscales associated with superconductivity are in the range of nanometres. Accordingly, measurements of electrical resistance over much larger distances are supposed to be insensitive to details of spatial inhomogeneities of electronic properties. We observe that, contrary to expectations, current paths adopt a highly non-uniform distribution at the onset of the superconducting transition which is manifested in the development of a finite transverse resistance. The anisotropic distribution of current density is unrelated to the structural properties of the superconducting films, and indicates the emergence of electronic inhomogeneities perceivable over macroscopic distances. Our experiments reveal the ubiquitous nature of this phenomenon in conventional superconductors.

\end{abstract}

                              %display desired
\maketitle

\newpage

%\pacs{Valid PACS appear here}% PACS, the Physics and Astronomy
                             % Classification Scheme.
%\keywords{Suggested keywords}%Use showkeys class option if keyword
                              %display desired
%\maketitle

%\tableofcontents

%\section{\label{sec:level1}First-level heading:\protect\\ The line
%break was forced \lowercase{via} \textbackslash\textbackslash}

\newpage

The typical lengthscales which govern spatial variations of superconducting order are the coherence length and the magnetic penetration depth. These parameters are of central importance for understanding the penetration of magnetic flux and the structure of the mixed state in type-II superconductors \cite{abrikosov,tinkham}. Electronic properties of mesoscopic samples are profoundly affected by the reduction in sample size when it reaches nanometric dimensions, comparable to the lengthscales characterizing fluctuations of the order parameter \cite{kanda, geim1, geim2, melnikov, schweigert, palacios}. In contrast, we do not expect the presence of microscopic inhomogeneities to be visible in measurements of macroscopic quantities, for example in the resistance of superconducting films when the distance between electrical probes is a few orders of magnitude larger than the coherence length. It is natural to treat the system as an isotropic conductor in this regime. In this work, we describe transport experiments in superconducting films of Nb and NbN which reveal that the impact of inhomogeneities in the electronic system is actually discernible in the form of a transverse resistance in the direction perpendicular to the macroscopic flow of current. Furthermore, the electronic inhomogeneities appear to be an emergent phenomenon, without any clear dependence on the morphology of the films. Our observations imply that the current density adopts non-uniform patterns, revealing unique aspects of emergent electronic properties which are not expected in such standard superconductors.

We will begin the discussion of our experiments by outlining the relation between anisotropy in a conductor and transverse resistance (Fig. 1). The resistance of a conducting film is characterized by applying an electrical current ($I$) and measuring the voltage drop ($V_{xx}$) along its path. The longitudinal resistance is $R_{xx}$ = $V_{xx}/I$. In a homogeneous planar conductor in the shape of a rectangle (Fig. 1a) with a current applied along its length, symmetry requirements determine that the lines of current density are parallel to each other. On the other hand, if we consider an inhomogeneous conductor (Fig. 1b) consisting of two different regions with resistivities $\rho_1$ and $\rho_2$ ($\rho_2 \ll \rho_1$), the current lines are attracted towards the region of lower resistivity leading to deviations from a symmetric distribution. The bending of current paths results in a transverse voltage drop $V_T$ between the points $A$ and $Z$ on opposite boundaries, in the direction perpendicular to the macroscopic flow of current $I$. The observation of a finite transverse resistance, $R_T=V_T/I$, implies the existence of macroscopic inhomogeneities.

\begin{figure}
\begin{center}
\includegraphics[width=88mm]{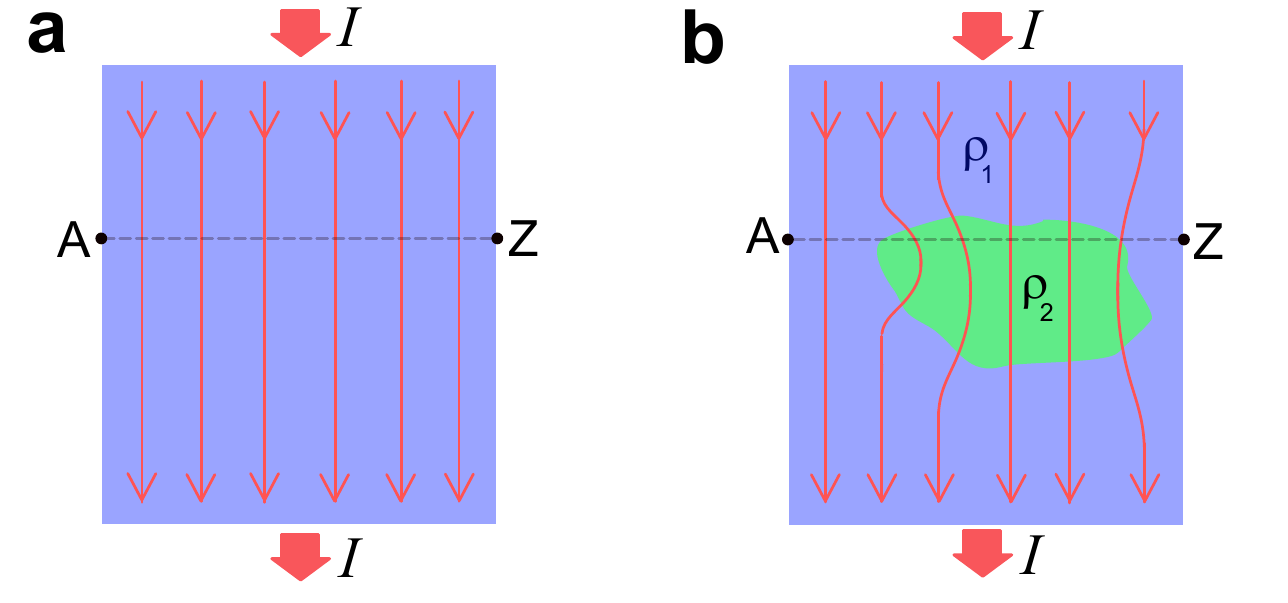}
\caption{\textbf{Current paths in inhomogeneous conductors.} \textbf{(a)} The lines of current density are uniformly distributed in an isotropic planar conductor. \textbf{(b)} In an inhomogeneous conductor with regions of two different resistivities $\rho_1$ and $\rho_2$ ($\rho_2 \ll \rho_1$), the current lines bend towards the region of lower resistivity. This leads to a voltage drop between the points $A$ and $Z$, in the direction perpendicular to the net flow of current.}
\end{center}
\end{figure}

\begin{figure*}
\begin{center}
\includegraphics[width=180mm]{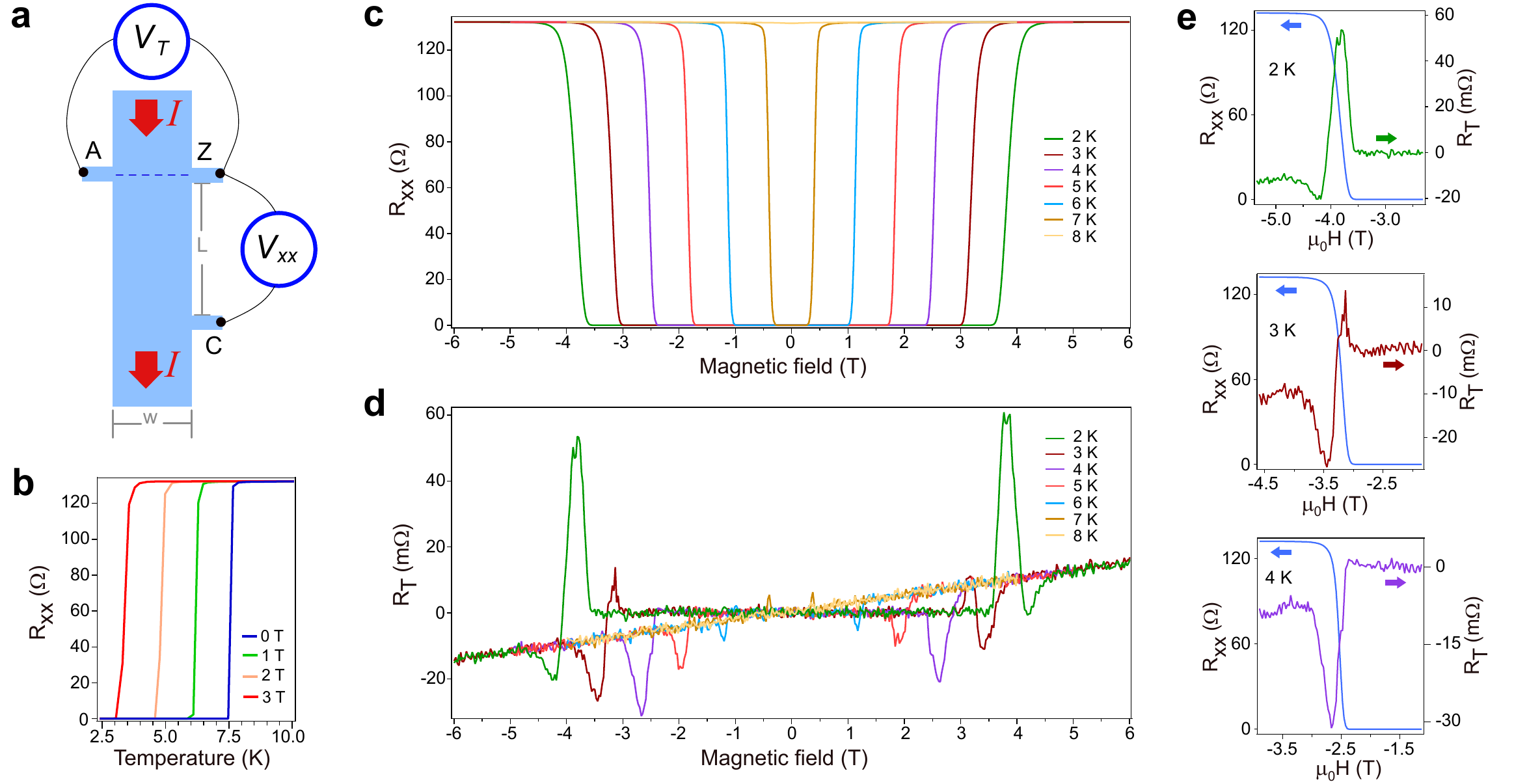}
\caption{\textbf{Transverse resistance in Hall bar devices of superconducting Nb.} \textbf{(a)} Schematic diagram of a Hall bar device for measuring $V_{xx}$ and $V_{T}$. \textbf{(b)} Measurement of $R_{xx}$ as a function of temperature at different values of the magnetic field to characterize the superconducting transition. An a.c. current of 25 $\mu$A r.m.s. amplitude was applied for the measurement. \textbf{(c,d)} $R_{xx}$ and $R_T$ were measured simultaneously as a function of magnetic field by applying an a.c. current of 50 $\mu$A r.m.s. amplitude. Peaks appeared in $R_T$ coinciding with the superconducting transition seen in $R_{xx}$ at the upper critical field. \textbf{(e)} Plots of $R_{xx}$ and $R_T$ from (c) and (d) are shown together for comparison, highlighting the variation in magnitude and sign of the peaks in $R_T$ at different temperatures. These variations signify the differences in flow patterns of electrical current near the critical field transition at different temperatures.}
\end{center}
\end{figure*}

We patterned devices on commercially available Si/SiO$_2$ substrates in the shape of Hall bars (Fig. 2a) using electron beam lithography, followed by the evaporation of Nb. The morphology of evaporated Nb films consist of amorphous and nanocrystalline regions\cite{durkin}. The electrical properties were measured using a standard low frequency lock-in technique. Fig. 2b shows the characterization of $R_{xx}$ as a function of temperature. In this device (named D1Nb), the voltage probes were a distance ($L$) 814 $\mu$m apart. The width ($w$) of the channel was 40 $\mu$m and its thickness ($t$) was 55 nm. The critical temperature ($T_c$) of the superconducting transition at zero magnetic field is 7.6 K. Figs. 2c and 2d show the result of simultaneous measurement of $R_{xx}$ and $R_T$ as a function of the magnetic field ($H$), at different values of temperature ($T$). The magnetic field was applied perpendicular to the plane of the sample in all our experiments. The upper critical magnetic field ($H_{c2}$) is determined from the midpoint of the resistance ($R_{xx}$) drop. It reaches upto 3.9 T at $T$ = 2.0 K. The Ginzburg-Landau coherence length is estimated\cite{helfand,werthamer} as $\xi_{GL}$ = [$\frac{\Phi_0}{2\pi H_{c2}\left(0\right)}$]$^\frac{1}{2}$, where $H_{c2}\left(0\right)$ = $0.69T_c\frac{dH_{c2}}{dT}|_{T=T_c}$ and $\Phi_0$ is the flux quantum. For this sample, $\xi_{GL}$ = 9.4 nm. At magnetic fields much higher than $H_{c2}$, the usual Hall effect is visible in the normal state leading to the linear variation of $R_T$ with the magnetic field. The estimated carrier density is 4.6$\times$10$^{22}$ cm$^{-3}$. The most interesting feature in the plot $R_T$($H$) is the occurrence of prominent even-in-field peaks (Figs. 2d and 2e) when the superconducting transition takes place.

Fig. 2d summarizes the most important point of this work, that there is a finite transverse resistance at the superconducting transition implying the current density is not uniform along the width of the channel between the transverse voltage probes. It can be inferred that there are inhomogeneous regions with a distribution of resistivity values inside the conductor, their impact being palpable even for large distances ($w$ = 40 $\mu$m) between the voltage probes. This distance is three orders of magnitude larger than $\xi_{GL}$. In order to check the robustness of this phenomenon, we did similar experiments on another conventional superconductor, NbN. This time, we studied an unpatterned plain NbN film with $t$ = 205 nm, deposited on MgO(011) substrate using DC magnetron sputtering. The film structure consists of nanosized cuboid domains (see Supplementary Information \cite{supp} for details on characterization of the sample). Electrical contacts were made by ultrasonic wire-bonding following the schematic in Fig. 3a. Resistance measurements were done by applying d.c. currents using a Quantum Design Model 6000 Physical Property Measurement System Controller. Current was injected between the contacts $I+$ and $I-$, 2.1 mm apart. The voltage $V_{xx}$ was measured between contacts $V+$ and $V-$, 0.7 mm apart. The transverse resistance $R_T$ was measured between the contacts $A$ and $Z$, 3.5 mm apart. The $T_c$ of this sample, named S1NbN, was 11.2 K (Fig. 3b). Figs. 3c and 3d show the results of $R_{xx}$ and ${R_T}$ measurement as a function of the magnetic field. $\xi_{GL}$ is estimated to be 2.7 nm from the $R_{xx}$($H$) data in Fig. 3c. In this system too, we observed (Fig. 3d) finite peaks in $R_T$ around the critical magnetic field, indicating that the distribution of current density is asymmetric with respect to the direction of macroscopic current flow.

Average values of the transverse ($\langle E_T\rangle$) and longitudinal ($\langle E_{xx}\rangle$) electric fields generated by an applied current are estimated from the measured voltages ($V_T$ and $V_{xx}$ respectively) divided by the distances between contacts. For sample D1Nb, $\langle E_T\rangle/\langle E_{xx}\rangle$ reaches a maximum of 2.9$\%$ near the upper critical field at $T$ = 2.0 K (Figs. 2c and 2d). In the case of sample S1NbN, the ratio reaches 5.9$\%$ at $T$ = 11.0 K (Figs. 3c and 3d).

An observation of great significance for the following discussion is that the features of the peaks in $R_T$($H$) plots are very different at different temperatures, as can be seen in Figs. 2d and 3d. The magnitude changes a lot, and often the sign as well. These features vary distinctly from one sample to another\cite{supp}, even for the same material (either Nb or NbN). We will now turn to the discussion of possible mechanisms of the described phenomenon.

Theoretical studies \cite{degennes,bardeen,nozieres} had anticipated the existence of novel effects in transport properties of superconductors due to the motion of vortices.  The Hall effect in many superconductors\cite{hagen1,hagen2,smith,khomskii} show distinctly different features at the onset of the transition compared to its normal state. However, these effects give rise to voltages with an antisymmetric dependence on the magnetic field and are thus unrelated to the even-in-field transverse resistance in our experiments.

Finite transverse resistances have been seen in certain superconducting films when a current was applied at zero magnetic field \cite{daluz,francavilla,antonova}. In some instances, an explanation based on vortex-antivortex interactions \cite{francavilla} had been suggested. This effect disappears when a small magnetic field \cite{antonova} is applied. Since the peaks in $R_T$ in our samples occur at large fields of few Teslas, such a mechanism based on vortex-antivortex interaction does not offer a suitable explanation.

\begin{figure}
\includegraphics[width=80mm]{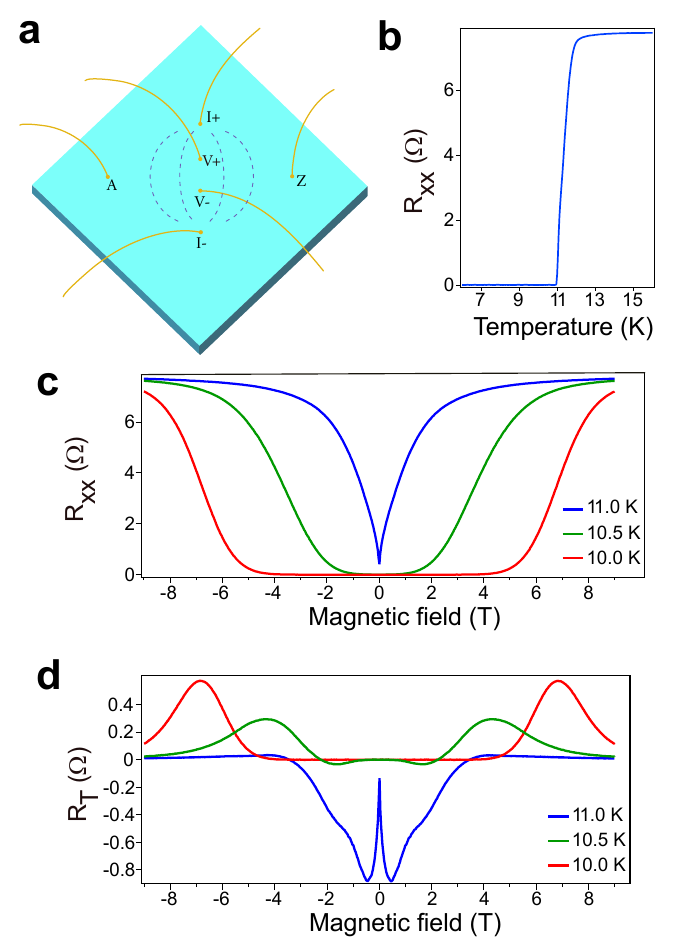}
\caption{\textbf{Transverse resistance in plain films of superconducting NbN.} \textbf{(a)} Electrical contacts are established by wire-bonding on the plain film. The current is injected between $I+$ and $I-$. The current density distributes along the plane (dashed lines). $V_{xx}$ is measured between $V+$ and $V-$. $V_T$ is measured between $A$ and $Z$. A d.c. current of 10 $\mu$A was applied. \textbf{(b)} Variation of $R_{xx}$ with temperature showing the superconducting transition. \textbf{(c,d)} $R_{xx}$ and $R_T$ were measured simultaneously as a function of magnetic field. A finite transverse resistance appears at the critical field transition. The peaks in $R_T$ do not have the same sign at all temperatures.}
\end{figure}

There are reports of superconductors showing transverse resistance peaks at the critical magnetic field\cite{segal, guryev} which are symmetric with respect to the field polarity. Segal et al. \cite{segal} provided a simple explanation for such even-in-field peaks based on the consideration that the superconducting film has non-uniform morphology with a spatial distribution of superconducting parameters ($T_c$ and $H_{c2}$). The analysis using a resistor network model shows that $R_T$ has a peak proportional in magnitude to $sgn(H)\frac{\partial R_{xx}(H)}{\partial H}$ when the transition is induced by varying the magnetic field. If it is induced by varying the temperature, $R_T$ shows a peak proportional to $\frac{\partial R_{xx}(T)}{\partial T}$. It provides a trivial explanation of even-in-field transverse resistance in terms of current paths guided by inhomogeneities ingrained in the sample structure. Experiments with magnetic imaging techniques have revealed the non-uniform distribution of electrical current in Nb networks\cite{wang}, which was attributed to disorder-driven variations of local $T_c$. Complex and non-uniform current flow patterns induced by defects were seen in polycrystalline TlBa$_2$Ca$_2$Cu$_3$O$_x$ superconducting films using magneto-optical imaging\cite{pashitski}. We will now see whether the aforementioned model based on structural inhomogeneities in the morphology of the film can explain our results. Let us consider the case of a non-uniform distribution of superconducting parameters across the width of a sample, for example, along the line joining probes $A$ and $Z$ in Fig. 2a. We assume that the right hand side (closer to $Z$) has on an average a slightly larger critical temperature and critical field. This should cause current lines to bend towards the right, leading to a positive voltage peak around $H_{c2}$ for a field-induced transition. This argument holds for all temperatures. Therefore, peaks in $R_T$($H$) should not differ much from one temperature to another, since the local superconducting properties are rooted in the structure of the film itself. They are expected to have the same sign, and also the lineshapes should not vary greatly due to its dependence on $sgn(H)\frac{\partial R_{xx}(H)}{\partial H}$. This is not what we observe in our experiments. The shapes of the $R_T$($H$) plots change drastically for different temperatures. More importantly, there is a change in sign of the peaks (Figs. 2d and 3d), despite the fact that $sgn(H)\frac{\partial R_{xx}(H)}{\partial H}$ is positive at all temperatures (Figs. 2c and 3c). Therefore, we find that the hypothesis of inhomogeneities in the material structure is unable to provide an explanation of the results, even qualitatively.

Xu et al. \cite{xu} recently reported the observation of anomalous even-in-field transverse resistance peaks in $\beta$-Bi$_2$Pd superconducting films. The transverse resistance was interpreted as a manifestation of chirality inherent in the system, and postulated to arise from topological surface states of $\beta$-Bi$_2$Pd. Such features have also been seen in the iron-based superconductor Ba$_{0.5}$K$_{0.5}$Fe$_2$As$_2$ \cite{lv}, possibly arising from an electronic nematic state. None of these explanations can be applied in our case since Nb and NbN films are not supposed to have either topological states or nematic ordering. It is the simplicity of these two systems which makes the observation of transverse resistance quite intriguing, suggesting that it reflects a phenomenon deeply rooted in the physics of superconducting criticality itself, irrespective of the structural and material properties.

Since the observation of transverse resistance in Nb and NbN films cannot be related to particular features of structure and morphology, it seems reasonable to conclude that the underlying phenomenon involves the development of electronic inhomogeneities of an emergent nature. The current density distribution, which determines the measured $R_T$, follows from the complex spatial variation of superconducting order in the vicinity of the phase transition. The essential points inferred from our experiments are the following. First, this electronic texture does not correspond to inhomogeneities ingrained in the film morphology. Second, it cannot be understood simply by taking into account order parameter variations over nanometric dimensions comparable to the coherence length, as these effects are expected to disappear upon macroscopic averaging and thus remain unnoticeable in resistance measurements. All these considerations prompt us to invoke the concept of emergence to describe the electronic inhomogeneities at the superconducting transition. Patterns of emergent behaviour in complex systems constitute fascinating topics of research in many branches of science \cite{bonabeau}. In solid-state physics, emergent properties are known in strongly correlated materials in which various types of physical interactions lead to the spontaneous development of electronic inhomogeneities at the nanometre scale\cite{dagotto} and novel electronic states\cite{hwang, yang}. Concerning the specific topic of superconductors, emergent behaviour has been seen \cite{kamlapure, carbillet} in disordered films close to the superconductor-insulator transition (SIT), characterized by nanosized domains much larger than typical lengthscales of structural inhomogeneities. There are very few cases where emergent behaviour leads to structures at macroscopic lengthscales. One example is the percolation network emerging in superconductor-graphene hybrid devices \cite{ioffe, allain} close to a SIT. In contrast, the devices studied by us are far from the limit of a SIT. The Ioffe-Regel parameter, defined as the product of Fermi wavevector ($k_F$) and mean free path ($l$), provides an estimate of disorder in a conductor and is expected to be close to 1 for samples near the SIT. The value of $k_Fl$ is 2.3 for the NbN sample S1NbN, and 30 for the Nb sample D1Nb. These samples cover a broad range of $k_Fl$ values extending well into the clean metallic regime for Nb, very far from a SIT. Therefore, the phenomenon of emergent electronic inhomogeneities spanning macroscopic distances, which we infer to be the reason behind the development of the transverse resistance, is unrelated to the physics of SIT. In this respect, the results reported here represent quite unique observations regarding the properties of superconductors.

The critical field transition in superconducting films is associated with the melting of the vortex lattice. The melting transition has been visualized at microscopic scales by scanning tunnelling microscopy in conventional superconducting films\cite{guillamon, indranil}. The transverse resistance observed in our samples is related to the flow pattern of electrical current in this regime. In order to obtain a deeper understanding of the phenomenon, it is necessary to develop a model concerning the statistical mechanics of large-scale pattern formation within the electronic system and the consequences on charge transport. The features of transverse resistance (seen in Figs. 2d and 3d) are reproducible even when the samples are heated and cooled across the critical temperature several times. This indicates that the patterns of macroscopic inhomogeneity are reproducible as a function of magnetic field and temperature, at least in a statistical sense. Two important factors influencing the melting transition are thermal fluctuations and the distribution of static quenched disorder in the superconducting films \cite{blatter}. It seems unlikely that these alone can explain the peaks in $R_T(H)$, which remain finite even upon macroscopic averaging and change signs at different temperatures. The non-trivial problem here is to understand what other factors might be relevant. This would require further research in the future.

We summarize the implications of this work for our conceptual understanding of the superconducting transition. The current density distribution in a homogeneous planar conductor is expected to be symmetric with respect to the direction of macroscopic current flow. We observe a finite transverse resistance at the onset of the superconducting transition, revealing that the condition of homogeneity is not respected. This indicates the presence of macroscopic inhomogeneities, palpable even when the voltage probes are separated by a distance at least few thousand times the characteristic Ginzburg-Landau coherence length. Measurements with a variation of both the temperature and magnetic field reveal that the inhomogeneities are not related to the material structure and morphology of the film, representing instead an emergent electronic phenomenon. We have demonstrated its occurrence in two different superconducting systems using two different contact geometries. Our work provides experimental evidence for a phenomenon that is unexpected in macroscopic samples. The central problem here concerns the nature of current flow patterns in superconductors and the lengthscales that are involved. An important goal for future research would be to arrive at an explanation for the occurrence of the transverse resistance at the critical field transition, as well as its strong dependence on temperature, from the perspective of microscopic theory. In recent times, imaging techniques have been used to map current flow patterns in a variety of systems, for probing the hydrodynamic transport of electrons in ultrapure conductors \cite{zeldov}, preferential paths created by domain structure at oxide interfaces\cite{kalisky2}, and fast vortices in superconductors\cite{embon} to give a few examples. Application of such techniques may provide complementary information regarding the transverse resistance observed in our transport experiments, and yield new insights about microscopic aspects of the phenomenon.

% The \nocite command causes all entries in a bibliography to be printed out
% whether or not they are actually referenced in the text. This is appropriate
% for the sample file to show the different styles of references, but authors
% most likely will not want to use it.
%\nocite{*}

%\bibliography{apssamp}% Produces the bibliography via BibTeX.

\begin{flushleft}

\newpage

%%%%%%%%%% Merge with supplemental materials %%%%%%%%%%
%\pagebreak
%\newpage
\widetext
\newpage
\begin{center}
\textbf{\large Supplementary Information}
\end{center}
%%%%%%%%%% Merge with supplemental materials %%%%%%%%%%
%%%%%%%%%% Prefix a "S" to all equations, figures, tables and reset the counter %%%%%%%%%%
\setcounter{equation}{0}
\setcounter{figure}{0}
\setcounter{table}{0}
\setcounter{page}{1}
\makeatletter
\renewcommand{\theequation}{S\arabic{equation}}
\renewcommand{\thefigure}{S\arabic{figure}}
\renewcommand{\bibnumfmt}[1]{[R#1]}
\renewcommand{\citenumfont}[1]{R#1}
%%%%%%%%%% Prefix a "S" to all equations, figures, tables and reset the counter %%%%%%%%%%

\bigskip

\begin{flushleft}

\bigskip

\textbf{1. Sample preparation and measurement setup}
\bigskip

For the fabrication of Nb devices, we used commercially available Si wafers with a 500 nm thick layer of insulating SiO$_2$ as the substrate. The devices were patterned using electron-beam lithography. We used resist bilayers of MMA/MAA (methyl methacrylate/methacrylic acid) and PMMA (polymethyl methacrylate). After the patterning of Hall bar devices with lithography, electron-beam-induced evaporation was performed at a pressure of typically 9$\times$10$^{-8}$ Torr with the sample holder at room temperature. The sample holder was rotating at 15 revolutions per minute. The lift-off process was completed by removing the resist layers with acetone. 

\bigskip

NbN thin films were prepared using an ultrahigh vacuum DC magnetron sputtering system with a base pressure of less than 4$\times$10$^{-10}$ Torr. Single crystals of MgO (011) with dimensions of 10 mm $\times$ 10 mm $\times$ 0.5 mm were used as substrates. The offcut of MgO (011) substrates was less than 0.5$^{\circ}$. The MgO substrates were cleaned following the method described in Ref. R1. They were then annealed in the deposition chamber for 30 minutes at 800$^{\circ}$C under ultrahigh vacuum before beginning the film growth. The NbN films were reactively sputter-deposited from a single-element, 3-inch Nb target (99.7$\%$ purity) at 800$^{\circ}$C with N$_2$ partial pressure of 0.6 mTorr. The working pressure was maintained at 4.5 mTorr. The substrate holder was rotating at 10 revolutions per minute. After deposition, the samples were cooled down to room temperature at a rate of 5$^{\circ}$C per minute.

\bigskip

Electrical contacts were established on the Nb and NbN samples using a wedge bonder. The samples were inserted in the cryostat of a Quantum Design Physical Property Measurement System (PPMS) for low temperature measurements. The electrical transport properties were measured using two different methods. The first method is a standard lock-in technique, carried out with Stanford Research Systems SR830 lock-in amplifiers. The second method involves the use of the Quantum Design Model 6000 PPMS Controller, which applies a d.c. current for determining the resistance. We used the first method for sample D1Nb, while the second one was used for sample S1NbN. Results from experiments on other samples of Nb and NbN using the lock-in technique are presented later.

\bigskip

\bigskip

\textbf{2. Analysis of experimental data on transverse resistance}

\bigskip

We have measured transverse resistance both in samples patterned in the shape of Hall bars and in plain films with electrical contacts established by wedge bonding. The underlying principle of all these measurements is that any deviation of current paths from a symmetric distribution (with the direction of macroscopic current flow being the axis of symmetry) may show up as a resistance $R_T$ in the transverse direction. Ideally, the line joining transverse voltage probes should be exactly perpendicular to the direction of applied current. However, this situation is most often not realized in practice since it is hard to avoid small misalignments of the voltage probes. Such misalignments cause a component of longitudinal resistance ($R_{xx}$) to appear in the measured signal. This has to be subtracted from the measured signal ($R_{T,m}$) of the transverse resistance. The estimate of $R_T$ is given by

\begin{equation}
R_{T} = R_{T,m} - (g*R_{xx})
\end{equation}

In the above equation, $g$ is a numerical factor characterizing the misalignment of transverse voltage probes. It is determined only by the position of the contacts, and has no dependence on temperature, applied magnetic field and current.

\bigskip

As we have discussed in the main article, if we assume that the deviation of current paths result mainly due to structural non-uniformities of the sample, the lineshape of transverse resistance as a function of the magnetic field is predicted\cite{segal2} to follow the function

\begin{equation}
\Lambda(H) \propto sgn(H)\frac{\partial R_{xx}(H)}{\partial H}
\end{equation}

The function $\Lambda(H)$ predicts the lineshape of transverse resistance, but not the sign of its peak. The sign may be either positive or negative depending upon the non-uniform distribution of superconducting parameters in the sample relative to the position of voltage probes. A consequence of this model is that $R_T(H)$ is predicted to have the same sign (either positive or negative) at all temperatures, since $R_{xx}(H)$ is a monotonic function around the superconducting critical field at any given temperature. We can compute $\Lambda(H)$ from the plots of $R_{xx}(H)$ measured at different values of temperature. For the samples discussed in the main article (D1Nb and S1NbN), we find that the observed nature of $R_T(H)$ is very different from $\Lambda(H)$. This discrepancy is the primary reason for interpreting the occurrence of the transverse resistance as a manifestation of emergent electronic inhomogeneities.

\bigskip

In a film of NbN with a thickness of 9 nm only, we have found that $R_T(H)$ indeed shows some similarities to $\Lambda(H)$. This data is presented in Section 4.3. Therefore, we note that the phenomenon of transverse resistance due to structural inhomogeneities reported by Segal et al.\cite{segal2} need not be entirely absent in our samples. However, any such effects, even if they are present, are overshadowed by a much larger contribution from electronic inhomogeneities as the film thickness becomes larger.

\bigskip

\begin{figure}[h!]
\begin{center}
\includegraphics[width=120mm]{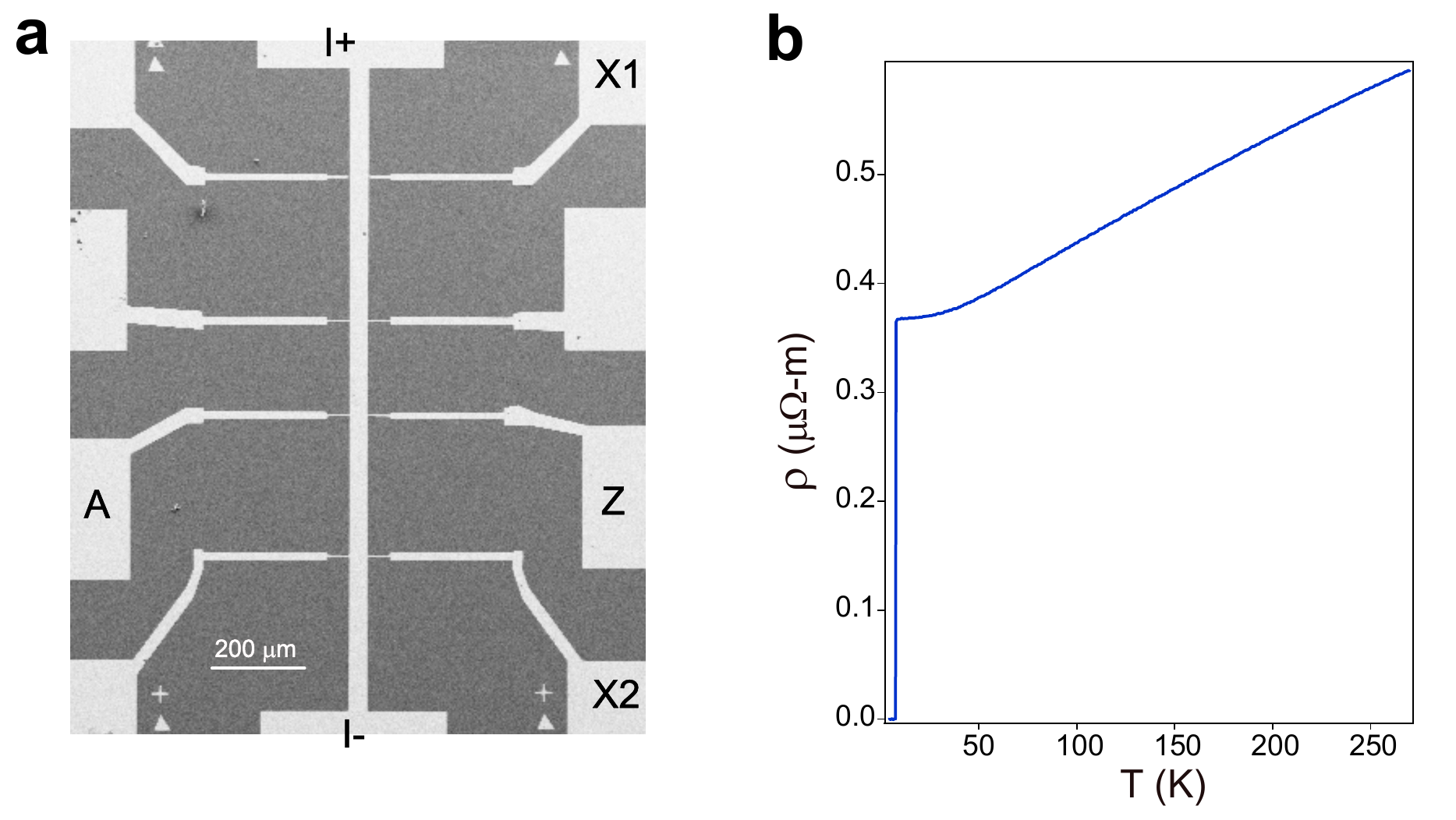}
\caption{\textbf{Resistivity measurement on sample D1Nb.} \textbf{(a)} Image of the device captured with a scanning electron microscope. \textbf{(b)} Resistivity ($\rho$) estimated from four-probe resistance measurements as a function of temperature ($T$). The current applied was 5 $\mu$A.}
\end{center}
\end{figure}

\begin{figure}
\begin{center}
\includegraphics[width=160mm]{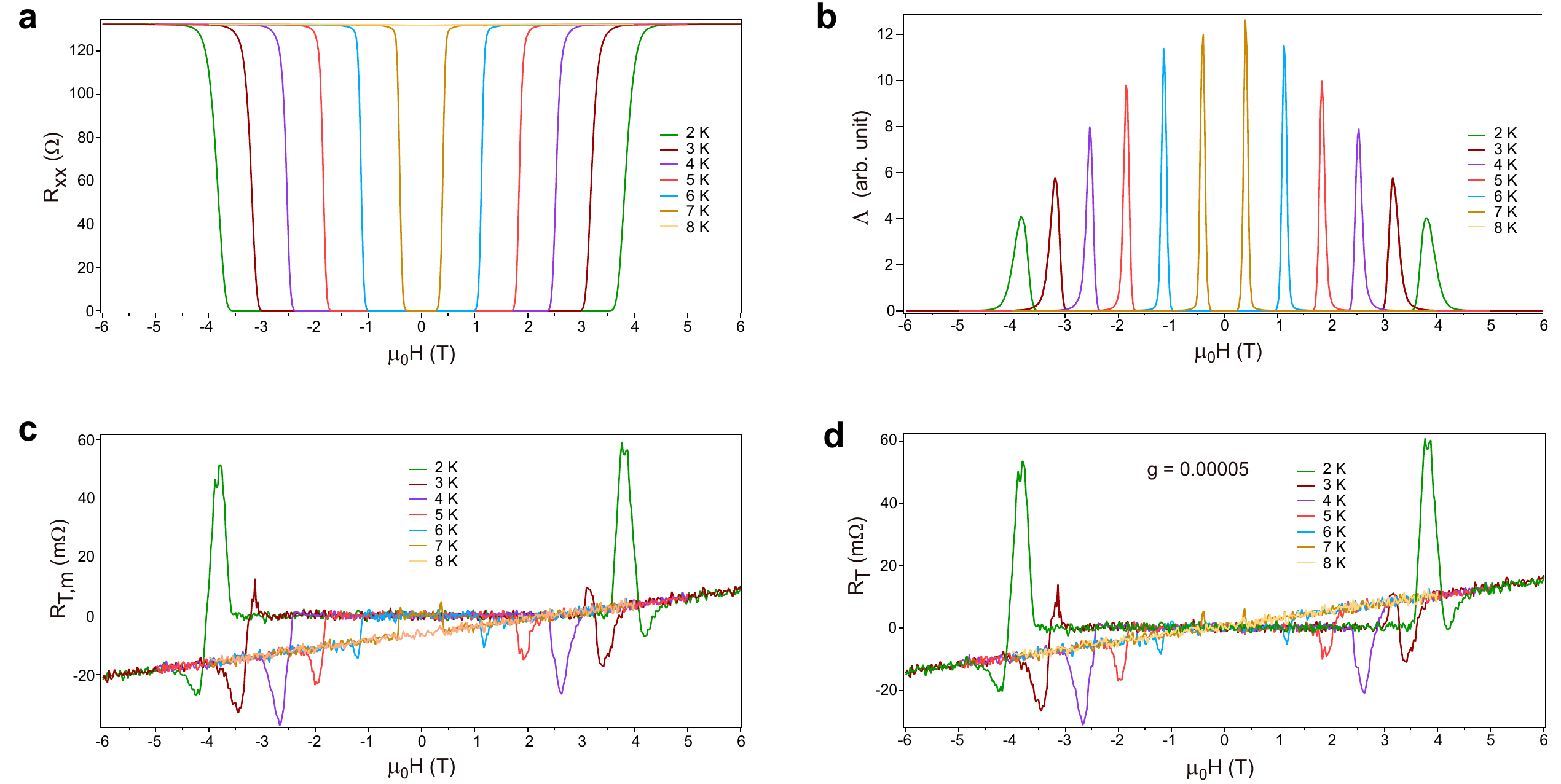}
\caption{\textbf{Analysis of transverse resistance measurement in sample D1Nb.} \textbf{(a)} The longitudinal resistance $R_{xx}$ was measured as a function of magnetic field ($H$) at different temperatures, using a low-frequency a.c. current of 50 $\mu$A. \textbf{(b)} The function $\Lambda$($H$) is plotted, following Eq. S2. \textbf{(c)} The raw data of transverse resistance measured between the probes $A$ and $Z$ (marked in Fig. S1a). \textbf{(d)} Plots of transverse resistance $R_{T}$ as a function of magnetic field, after the background signal proportional to $R_{xx}(H)$ is subtracted from the $R_{T,m}(H)$ plots following Eq. S1, with $g$=0.00005. This figure is the same as Fig. 2d in the main article.}
\end{center}
\end{figure}

\begin{figure}
\begin{center}
\includegraphics[width=160mm]{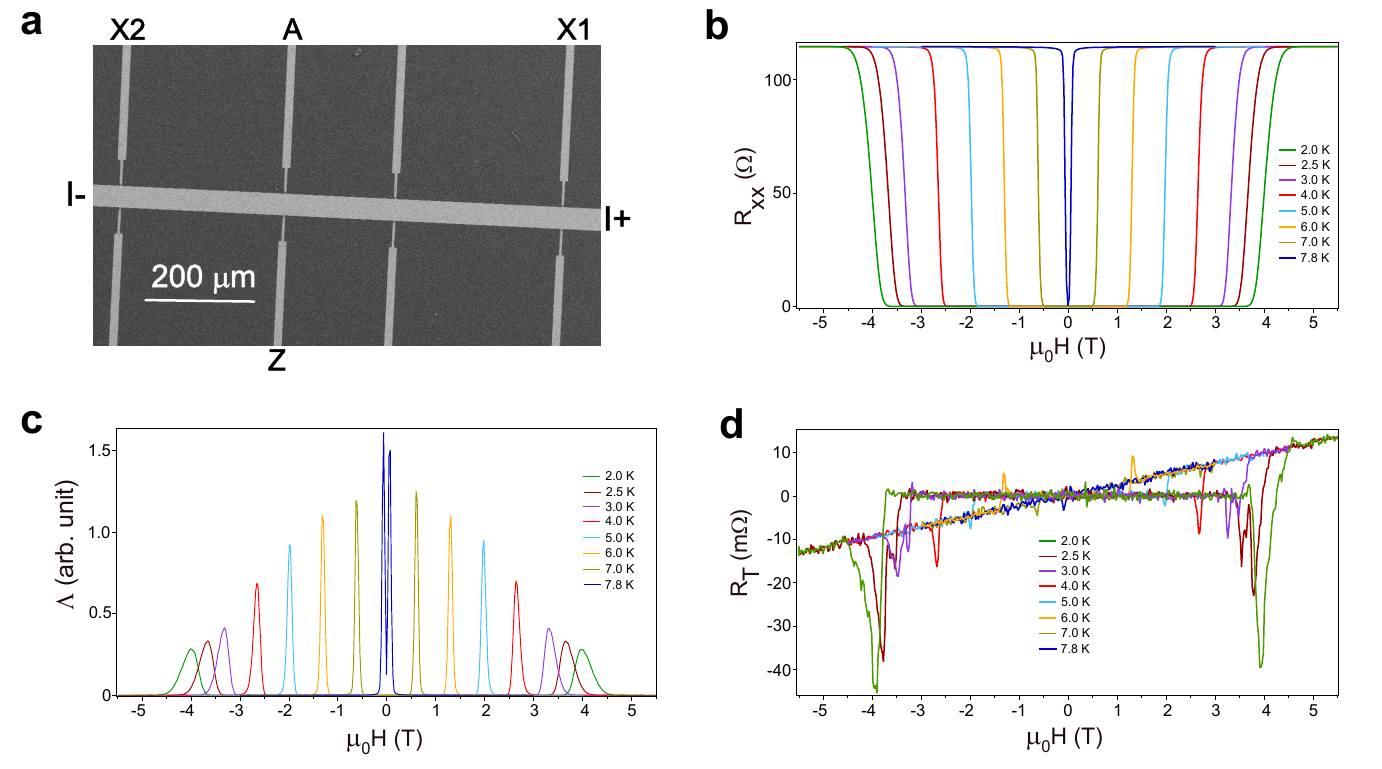}
\caption{\textbf{Measurement of transverse resistance in sample D2Nb.} \textbf{(a)} Image of the device captured with a scanning electron microscope. Its thickness is 55 nm. \textbf{(b)} The longitudinal resistance ($R_{xx}$), measured between the probes $X1$ and $X2$, as a function of magnetic field ($H$) at different temperatures. A low-frequency a.c. current of 50 $\mu$A was applied for the measurement. \textbf{(c)} The function $\Lambda(H)$ is estimated from the $R_{xx}(H)$ data. \textbf{(d)} Plots of transverse resistance (measured between contacts A and Z) as a function of magnetic field.}
\end{center}
\end{figure}

\textbf{3. Experiments on Nb samples}

\bigskip

We present below the results of experiments on different samples of superconducting Nb films. We will begin with the sample D1Nb, which has aready been discussed in the main article. The procedure of analyzing the raw data will be described in detail. Results from two other samples (D2Nb and D3Nb) will also be presented. All the resistance measurements were conducted by a low-frequency lock-in technique.

\bigskip

\textbf{3.1 Sample D1Nb}

\bigskip

An image of the device is shown in Fig. S1a. The variation of resistivity ($\rho$) of this sample as a function of temperature ($T$) is shown in Fig. S1b.

\bigskip

Fig. S2 displays the same data presented in Figs. 2c and 2d of the main article. We will now describe how $R_T$ is estimated following Eq. S1. For these experiments, current was applied between the contacts $I+$ and $I-$ (Fig. S1a). Longitudinal resistance ($R_{xx}$) was measured between $X1$ and $X2$, while the transverse resistance ($R_{T,m}$) was measured between $A$ and $Z$. The variation of $R_{xx}$ as a function of magnetic field ($H$) is shown in Fig. S2a. These plots are used to estimate the critical field $H_{c2}$ at different temperatures and the Ginzburg-Landau coherence length $\xi_{GL}$. We obtained $\xi_{GL}$ = 9.4 nm. As we have discussed earlier, if we take into account the deviation of current paths owing to structural inhomogeneties of the sample only, the lineshape of transverse resistance $R_T$ is predicted to follow the function $\Lambda(H)$, which is shown in Fig. S2b.

\bigskip

The raw data of measured transverse resistance $R_{T,m}(H)$ is presented in Fig. S2c. Since the probes of the Hall bar device are not perfectly aligned, there is a small contribution of the longitudinal voltage drop (proportional to $R_{xx}$) which appears in $R_{T,m}(H)$. This becomes evident if we look at the high-field behaviour when the Hall effect in normal state is visible (Fig. S2c). The Hall voltage is an odd function of $H$. But we see that the magnitudes of the transverse resistance are not exactly equal at positive and negative values of the magnetic field. The background signal proportional to $R_{xx}(H)$ is subtracted from the $R_{T,m}(H)$ plots following Eq. S1, with $g$=0.00005. The resulting plots of $R_T(H)$ are shown in Fig. S2d (which is identical to Fig. 2d of the main article). It is evident that the peaks in $R_T(H)$ do not follow the trend of $\Lambda(H)$.

\bigskip

From the Hall resistance data at high magnetic fields, the mean free path ($l$) is estimated to be 2.8 nm using the free electron model. This is an approximate estimate, given the fact that niobium has a complicated Fermi surface \cite{beaulac} with positive Hall coefficient.

\begin{figure}
\begin{center}
\includegraphics[width=80mm]{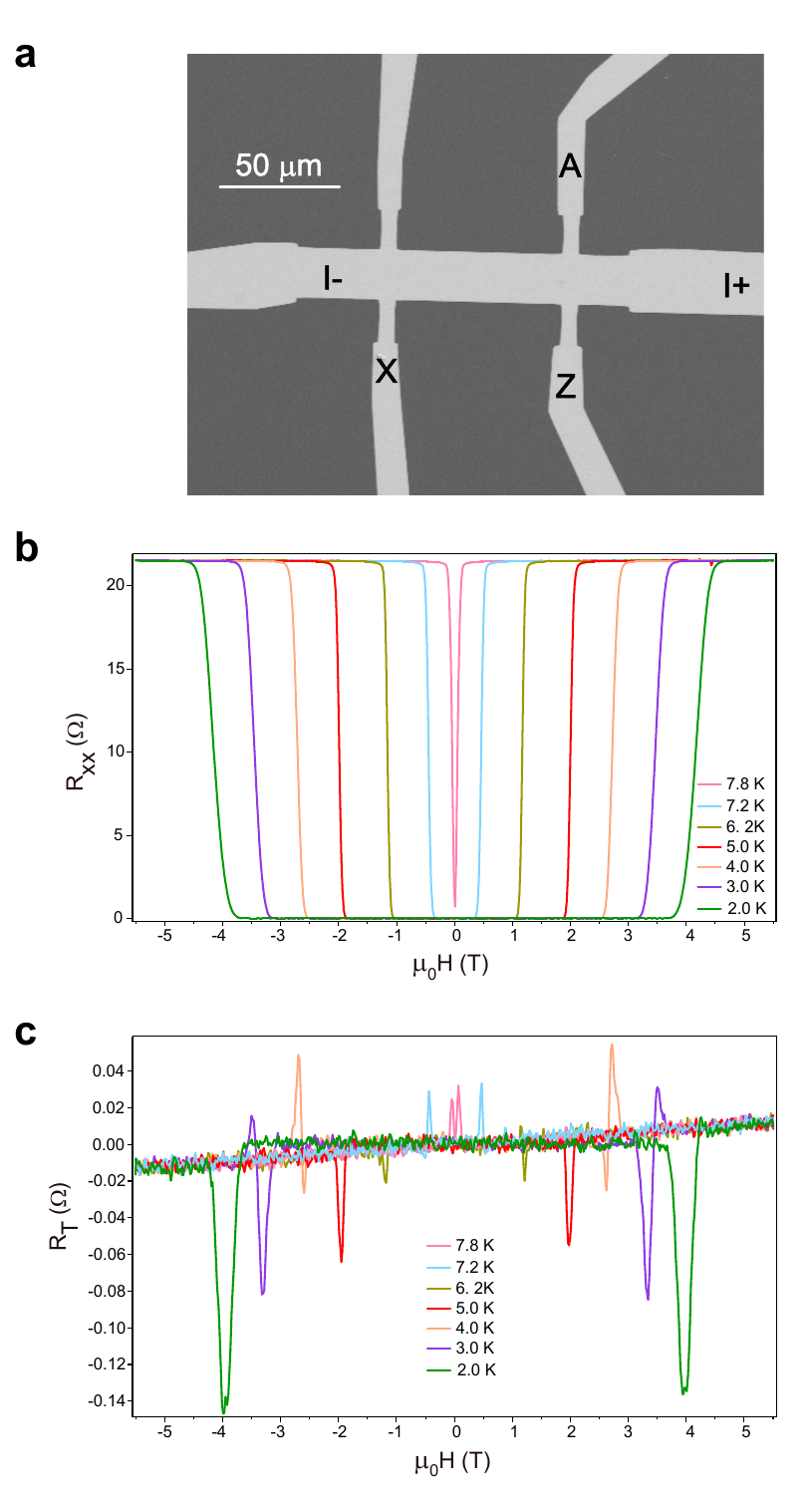}
\caption{\textbf{Measurement of transverse resistance in sample D3Nb.} \textbf{(a)} An SEM image of the device. \textbf{(b)} The longitudinal resistance ($R_{xx}$), measured between the probes $Z$ and $X$, as a function of magnetic field ($H$) at different temperatures. A low-frequency a.c. current of 5 $\mu$A was applied for the measurement. \textbf{(c)} The transverse resistance (measured between A and Z) is shown as a function of magnetic field at different temperatures.}
\end{center}
\end{figure}

\bigskip

\bigskip

\textbf{3.2 Sample D2Nb}

\bigskip

Sample D2Nb was prepared with a similar design and under identical conditions as sample D1Nb. Fig. S3a shows an image of the device. Current was applied between $I+$ and $I-$. The variation of four-probe resistance ($R_{xx}$) as a function of magnetic field ($H$) is shown in Fig. S3b. 

\bigskip

Fig. S3c plots $\Lambda(H)$ at different values of temperature using the data of $R_{xx}(H)$. The transverse resistance observed at the critical field transition is shown in Fig. S3d. The peaks in $R_T(H)$ in this case are found to have generally the same sign at all temperatures, except for $T$ = 6.0 K. The peaks become more prominent at low temperatures, which contrasts with the predicted trend in Fig. S3c.

\bigskip

\bigskip

\bigskip

\textbf{3.3 Sample D3Nb}

\bigskip

Sample D3Nb is a Hall bar device of Nb with a narrower channel than the samples previously discussed. It has a width on 20 $\mu$m. The thickness is 60 nm. An SEM image is shown in Fig. S4a. The critical temperature ($T_c$) of this device is 7.9 K.

\bigskip

The variation of $R_{xx}$ as a function of magnetic field is shown in Fig. S4b. Fig. S4c shows the results of $R_T(H)$ measurements. This sample exhibits multiple reversals of the sign of $R_T$ peaks. Starting from low temperature ($T$ = 2.0 K) with a negative peak, one reversal to positive sign occurs at $T$ = 4.0 K. It becomes negative again at $T$ = 5.0 K. A further reversal to positive is seen at $T$ = 7.2 K. The current density distribution across the width of the channel (between the probes $A$ and $Z$) seems to vary in an erratic manner as the temperature is changed.

\bigskip

\bigskip

\bigskip

\bigskip

\bigskip

\textbf{4. Experiments on NbN samples}

\bigskip

The NbN samples were deposited as plain films on sapphire and MgO substrates. The resistance measurements were conducted following the schematic outlined in Fig. 3a of the main article. Experiments on the sample S1NbN have been presented in the main article. We discuss below certain aspects of this sample, as well as results from two other samples.

\begin{figure}[h!]
\begin{center}
\includegraphics[width=140mm]{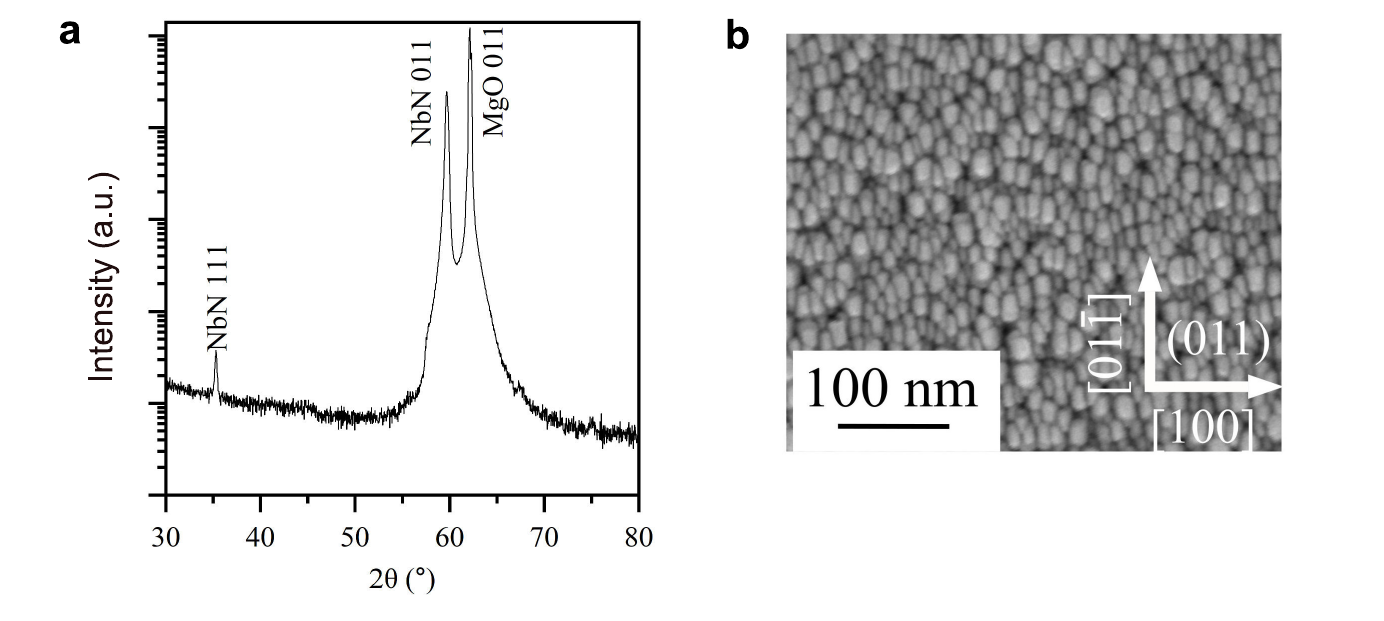}
\caption{\textbf{Morphology of NbN film deposited on MgO(011).} \textbf{(a)} Result of X-ray diffraction (XRD) experiment. \textbf{(b)} A top-view scanning electron micrograph.}
\end{center}
\end{figure}

\bigskip

\bigskip

\bigskip

\textbf{4.1 Sample S1NbN}

\bigskip

The NbN film S1NbN was prepared on a MgO(011) surface. The deposition procedure was described in Section 1. Fig. S5 shows the morphology of a NbN film on MgO(011) substrate. $\theta$-2$\theta$ X-ray diffraction (XRD) patterns were obtained (Fig. S5a) using a PANalytical X'Pert PRO diffractometer with Cu-K$\alpha$ radiation ($\lambda$ = 1.54060 $\AA$) at 45 kV and 40 mA. The scans were carried out in the Bragg-Brentano configuration. Cu-K$\beta$ was removed using a Ni filter. Fixed slits of 1/2$^{\circ}$ were used as anti-scatter and divergence slits. The step size of $\theta$ for this measurement was 0.03$^{\circ}$. Notably, a minor NbN 111 peak can only be seen on a logarithmic scale. The dominance of the NbN 011 peak over NbN 111 by over 100-fold suggests a limited percentage of randomly oriented NbN crystals. Fig. S5b shows an image of a film captured with an SEM. The microsctructure consists of nanoscale cuboid domains. The arrangement of NbN cuboids is in accordance with the in-plane orientation of MgO (011) as depicted in the figure.

%The resistance of sample S1NbN (Fig. S5a) showed negative temperature coefficient (i.e., $\frac{dR_{xx}}{dT}<0$) for a wide range of temperature before the superconducting transition at 11.2 K. The insulating behaviour at high temperatures is associated with low values of the Ioffe-Regel parameter $k_Fl$ in NbN films. The resistivity of the film at 15 K is 7.0 $\mu\Omega$-m.

\bigskip

\begin{figure}[h!]
\begin{center}
\includegraphics[width=160mm]{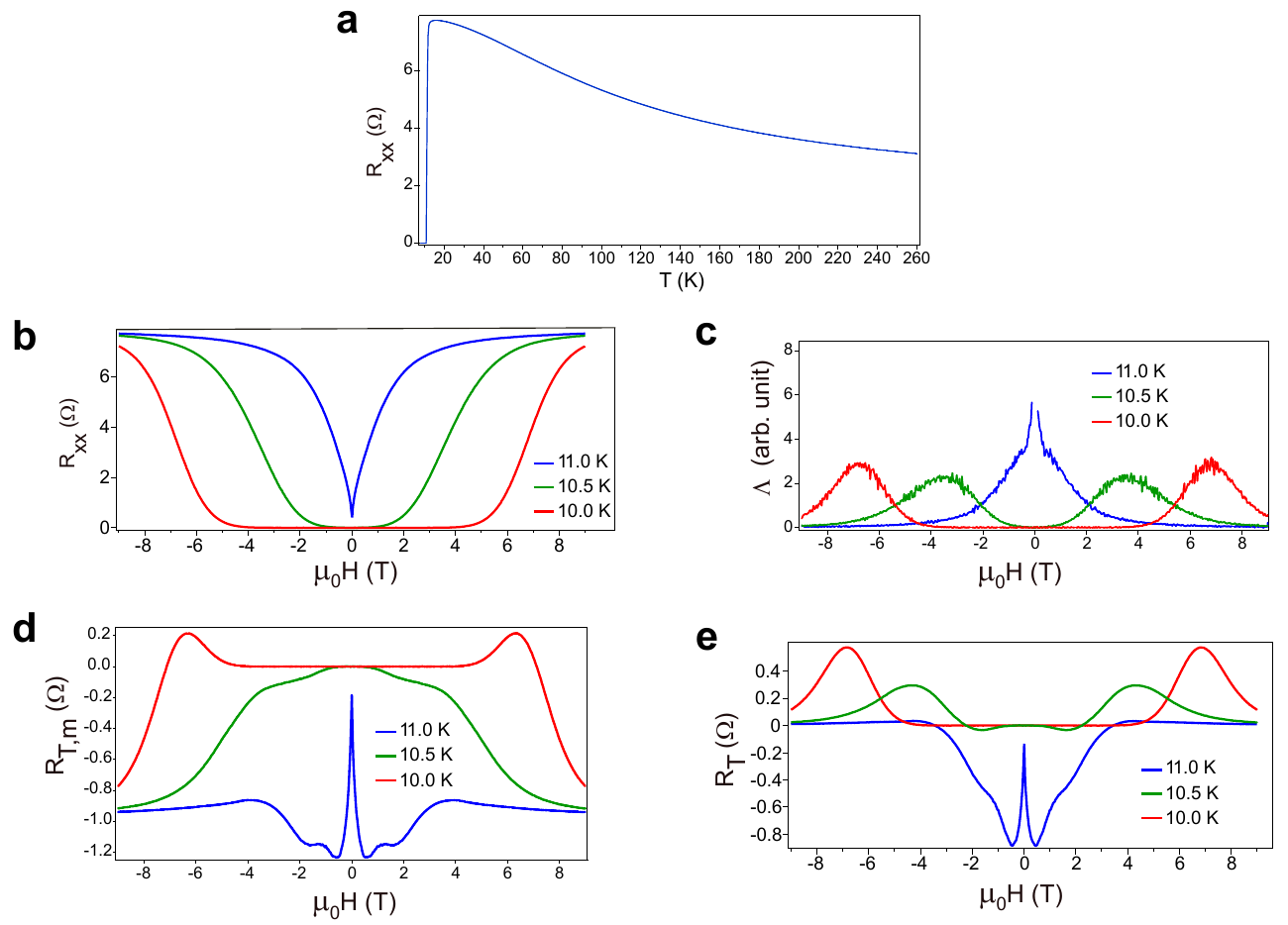}
\caption{\textbf{Resistance measurements on sample S1NbN.} \textbf{(a)} The resistance $R_{xx}$ measured as a function of temperature. The superconducting transition occurs at 11.2 K. These measurements were done using the bridge of a Quantum Design Model 6000 Physical Property Measurement System. A d.c. current of 40 $\mu$A was applied. \textbf{(b)} The resistance $R_{xx}$ measured as a function of magnetic field using a d.c. current of 10 $\mu$A. \textbf{(c)} The function $\Lambda$($H$) is plotted. \textbf{(d)} The raw data of transverse resistance measured between the transverse voltage probes. \textbf{(e)} Plots of transverse resistance as a function of magnetic field after subtracting a background proportional to $R_{xx}$. The parameter $g$, used for the analysis in Eq. 1, is 0.1235. This figure is the same as Fig. 3d in the main article.}
\end{center}
\end{figure}

Resistance measurements on the sample S1NbN over a large range of temperatures are shown in Fig. S6a. The measurements were conducted using the Quantum Design Model 6000 PPMS Controller, which applies a d.c. current for measuring the resistances. The resistivity of the film at 15 K is 7.0 $\mu\Omega$-m. We will now describe the procedure of analyzing the data concerning the transverse resistance on this sample.  The plots of $R_{xx}$ as a function of magnetic field are shown in Fig. S6b (identical to Fig. 3c of the main article). These plots are used to estimate $\Lambda(H)$ in Fig. S6c. The measured data for $R_{T,m}(H)$ is shown in Fig. S6d. There is a contribution of longitudinal voltage drop, because of the misalignment of the transverse resistance probes. $R_T$ is estimated by correcting for this contribution using Eq. S1, with g=0.1235. These plots are shown in Fig. S6e, which is the same as Fig. 3d of the main article.

\bigskip

Resistance measurements reveal very broad transitions at the critical magnetic field (Fig. S6b) with extremely large value of $\left[\frac{d(\mu_0H_{c2})}{dT}\right] _{T=T_c}$ = 6.0 T/K. We obtain $\mu_0H_{c2}(0)$ = 46 T and $\xi_{GL}$ = 2.7 nm. Such large values of $H_{c2}(0)$ in NbN films have been reported earlier and attributed to the presence of small column void microstructures\cite{ashkin}. In our sample, the high critical field is probably related to the occurrence of nanoscale cuboidal domains (Fig. S5b).

\bigskip

\begin{figure}
\begin{center}
\includegraphics[width=140mm]{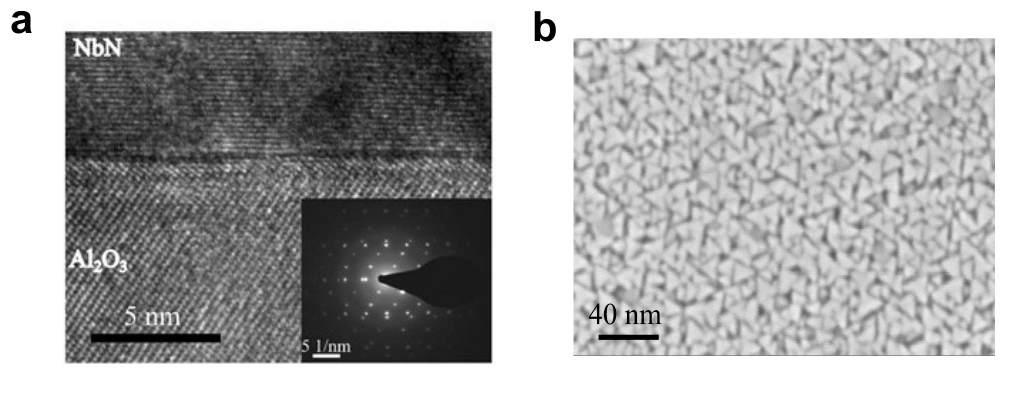}
\caption{\textbf{Structure of NbN films on Al$_2$O$_3$ (0006) substrate.} \textbf{(a)} Cross-sectional TEM image reveals the detailed interface structure between NbN and sapphire. The inset presents the corresponding selected area diffraction pattern. \textbf{(b)} Scanning electron micrograph shows triangular domains.}
\end{center}
\end{figure}

\begin{figure}
\begin{center}
\includegraphics[width=100mm]{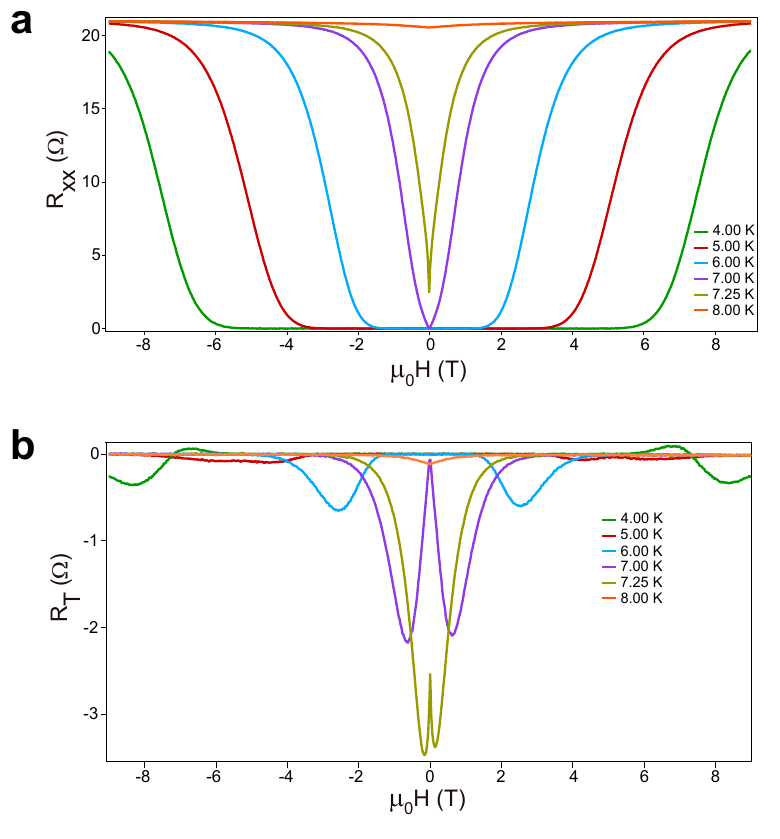}
\caption{\textbf{Experiments with Sample S2NbN.} \textbf{(a)} The resistance $R_{xx}$ measured as a function of magnetic field using a lock-in technique. An a.c. current of 10 $\mu$A r.m.s. was applied. \textbf{(b)} Plots of transverse resistance as a function of magnetic field.}
\end{center}
\end{figure}

\bigskip

\bigskip

\textbf{4.2 Sample S2NbN}

\bigskip

Sample S2NbN is a NbN film with a thickness of 60 nm. It was prepared on single crystal c-plane sapphire (0006) under similar conditions as those used for growth on MgO with a slight difference in the cleaning procedure. In this case, the substrates underwent ultrasonic cleaning in acetone and ethanol for 10 minutes, prior to loading into the chamber.

\bigskip

The morphology of the films on sapphire was examined using transmission electron microscopy (TEM) utilizing an FEI Tecnai G2 TF 20 UT microscope operating at 200 kV. Cross-sectional samples for TEM were first manually polished to approximately 60 $\mu$m thickness and then subjected to Ar+ ion milling at 5 keV with a 5$^\circ$ tilt, while being rotated in a Gatan precision ion polishing system. Fig. S7a represents the high-resolution TEM image of NbN grown on c-plane sapphire. The interface between NbN and sapphire appears to be sharp and well-defined, indicating the absence of an interdiffusion layer. The NbN layer is relatively uniform and crystalline, as shown by the regular pattern of lattice fringes visible throughout the layer. However, the fringes are less distinct compared to the high-quality sapphire, indicating the possible presence of some domains.

\bigskip

Fig. S7b shows the top-view SEM image of a typical epitaxial NbN film grown on c-plane sapphire. The film consists of triangular domains caused by different symmetries of sapphire and NbN\cite{iovan}. Based on this figure, the sizes of the domains are mostly in the range of 5 to 10 nm which can be tuned by varying the deposition conditions. However, the overall morphology of triangular domains remains consistent. 

\bigskip

The sample S2NbN has a $T_c$ of 7.5 K. The results of resistance measurement as a function of magnetic field are presented in Fig. S8a. The normal state resistance is 21 $\Omega$, corresponding to a resistivity of 3.1 $\mu\Omega$-m.

\bigskip

The peaks observed in $R_T(H)$ (Fig. S8b) at the critical field transition reduce considerably in magnitude as the temperature is lowered from 7.25 K to 6.00 K. There is an anomalous behaviour at $T$ = 5.00 K. Here the $R_T(H)$ curve becomes very flat with a low value, indicating that the current density is more evenly distributed across the film at $T$ = 5.00 K, compared to other temperatures. Upon lowering the temperature further to $T$ = 4.00 K, the peak becomes larger in magnitude again.

\begin{figure}
\begin{center}
\includegraphics[width=120mm]{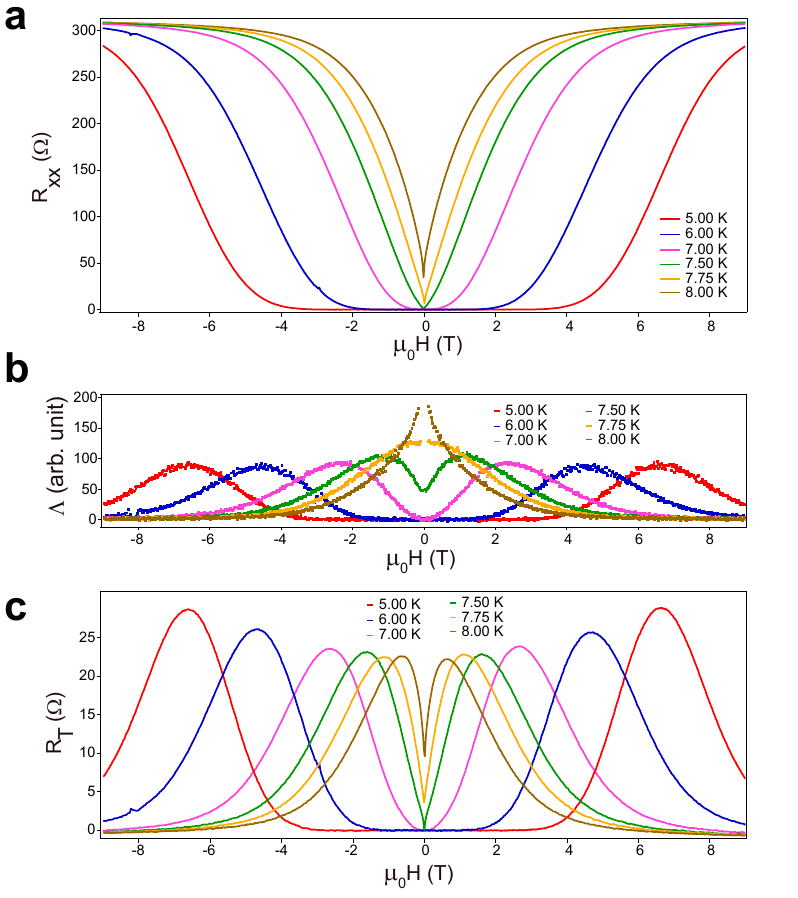}
\caption{\textbf{Experiments with Sample S3NbN.} \textbf{(a)} The resistance $R_{xx}$ measured as a function of magnetic field using a lock-in technique. A current of 1 $\mu$A r.m.s. amplitude was applied. \textbf{(b)} $\Lambda(H)$ is plotted . \textbf{(c)} Plots of transverse resistance as a function of magnetic field.}
\end{center}
\end{figure}

\bigskip

\textbf{4.3 Sample S3NbN}

\bigskip

Sample S3NbN is a film of NbN deposited on MgO(011) substrate. It has a thickness of 9 nm and is the thinnest of all films measured by us. The critical temperature was found to be 8.5 K. The critical field measurements are shown in Fig. S9a. The normal state resistance measured in four-probe configuration is 309 $\Omega$. This corresponds to a resistivity of 9.9 $\mu\Omega$-m.

\bigskip

The transverse resistance measured in this film offers important insights about the phenomenon. The expected variation of $R_T(H)$, following the lineshape given by $\Lambda(H)$, is depicted in Fig. S9b. Unlike the previously reported samples, we see here that the measured $R_T$ (shown in Fig. S9c) actually follows the predicted trend of $\Lambda(H)$ to a fair degree. The lineshapes of experimentally observed $R_T(H)$ (in Fig. S9c) are quite similar at all temperatures. Although the magnitude varies a bit at different temperatures, the difference is not very large. Thus, we see here a behaviour which conforms well to the prediction of the model of Segal et al.\cite{segal2}. In other words, the impact of inhomogeneous distribution of superconducting parameters intrinsic in the structure of the film is the dominant factor for the transverse resistance in this sample. This provides clearly a different case than all the other samples measured by us. Since this is also the thinnest sample with $t$ = 9 nm, we conclude that the impact of emergent electronic inhomogeneities on the development of the transverse resistance disappears in the limit of small film thickness. In this regime the structural non-uniformities play the dominant role in guiding the paths of electric current.

\end{flushleft}

\end{flushleft}

\makeatletter
\renewcommand\@biblabel[1]{[R#1]}
\makeatother

\end{document}